\documentclass[12pt]{article}
\usepackage{epsfig}

\newcommand{\bn}{\bar\nu}

\newcommand{\oot}{\overline {126}}

\newcommand{\nnu}{\nonumber\\}
\newcommand{\be}{\begin{equation}}
\newcommand{\ee}{\end{equation}}
\newcommand{\bea}{\begin{eqnarray}}
\newcommand{\eea}{\end{eqnarray}}

\begin{document}

 \vfil
 \vspace{3.5 cm} \Large{
 \title{\bf  {  MSGUT Reborn ? }}
 \author{ Charanjit S. Aulakh   }}
\date{}
\maketitle

 \normalsize\baselineskip=15pt
{\centerline  {\it High Energy Theory Group, ICTP}} {\centerline{
\it {Trieste, Italy 34100}}}

{\centerline  {\it and }}

 {\centerline  {\it
Dept. of Physics, Panjab University}} {\centerline{ \it
{Chandigarh, India 160014}}}

 {\centerline {\rm E-mail: aulakh@pu.ac.in }}
\vspace{1.5 cm}

\large {\centerline{\bf {ABSTRACT }}}
\normalsize\baselineskip=15pt

\vspace{1. cm}    We present   examples of fits of fermion mass
data using the $\mathbf{10-120-\oot}$ FM Higgs system in Susy
SO(10) GUTs that follow the  scenario\cite{blmdoom,nmsgutI}  in
which the $\mathbf{120}$ -plet collaborates with the $
\mathbf{10}$-plet to fit the charged fermion masses while  small
$\mathbf{\oot} $-plet couplings enhance the Type I seesaw
neutrino  masses   to viable values and  make the fit to light
fermions accurate. Restricting ourselves to the CP conserving case
we use a linear perturbative technique to obtain accurate charged
fermion fits ($\chi^2_{tot} < .2 $) to all the charged fermion
masses and angles. The resulting fits imply Type I neutrino masses
that are generically $10^2- 10^3$ times larger than those obtained
in the $\mathbf{10-\oot}$ scenario precisely because of the small
$\mathbf{\oot} $ coupling. Thus the difficulty of obtaining
sufficiently large neutrino masses in the context of these  Next to Minimal
Susy GUTs is  essentially removed. The remaining free parameters
 allow one to obtain the correct ratio of neutrino mass squared splitting and a
large($\theta_{23}^{PMNS}$) and small($\theta_{13}^{PMNS}$) mixing
angle. The $\theta_{12}^{PMNS}$ is however small and this
indicates that -as in the Type I $\mathbf{10-\oot}$ case - a fully
realistic  fit to the lepton mixing data also  \emph{ requires} CP
violation.

\section{Introduction}

The discovery of neutrino oscillations in 1998, followed by the
measurement of  increasingly accurate values of the neutrino mass
squared differences and mixing angles has driven an intense wave
of research into supersymmetric seesaw mechanisms, Left-right
symmetric models and particularly all aspects of the structure of
renormalizable Susy SO(10) GUTs.  In particular the minimal
Supersymmetric  Grand Unified Theory (MSGUT)
\cite{aulmoh,ckn,abmsv}  namely a Supersymmetric SO(10) GUT based
on a $\mathbf{{210-10-\oot-126}}$  Higgs system has received a
great deal of attention over the last seven years. Detailed
calculations of every aspect of its
spectra\cite{ag1,bmsv,fuku04,ag2}, couplings\cite{ag1,ag2} and
their evolution\cite{ag1,ag2} have been published by several
groups. Besides the question of gauge unification the central
problem of the second phase of the Grand Unification program
namely that of accounting for the observed fermion mass spectrum
and mixing structure, for both charged fermions and neutrinos,
using the mass relations dictated by the simplest possibly
realistic $\mathbf{10-\oot}$ Higgs structure \cite{babmoh}, has
inspired a great deal of detailed analysis of the fitting
problem\cite{bsv,gohmoh,bert,babmacesanu}.

This so called Babu-Mohapatra(BM) program had, until last year,
met with   uniform success in accounting for the observed charged
fermion spectrum, neutrino mass splitting  and mixing using the
\emph{generic} structure of the fermion mass matrices dictated by
the choice of Higgs ($\mathbf{10-\oot}$) and the Type Ib and Type
II \cite{seesaw}  Seesaw mechanisms naturally present in SO(10)
models. Successful fits of all the fermion mass data available
including the neutrino mixing data \cite{strumviss} using Type I ,
Type II and mixed  seesaw mechanisms were
obtained\cite{matsu0,bsv,gohmoh,bert,babmacesanu}. It was,
however, just  assumed that the overall scale and the relative
magnitudes of the Type I and Type II seesaw masses required for
the fits would be realizable in appropriate Susy GUTs. The only
such available theory, where in fact the calculations of
spectra\cite{ag1,ag2,bmsv,fuku04} and
couplings\cite{ag1,ag2,gmblm} necessary to perform precisely such
a test had almost contemporaneously been completed, was the MSGUT
itself.  The very first survey\cite{gmblm} revealed that there
were serious difficulties in obtaining Type II over Type I
dominance as well as in obtaining large enough Type I neutrino
masses. Using a convenient parametrization of the  MSGUT spectra
and couplings in terms of the  single ``fast''   parameter ($x$)
which controlled MSGUT symmetry breaking
\cite{abmsv,bmsv,ag2,bmsv2} a complete proof was then
given\cite{blmdoom} of the failure of the Seesaw mechanism in the
context of the MSGUT.  The nature of the obstruction uncovered by
us also led us to suggest a very natural scenario that might deal
with the problem in the MSGUT extended by adding  the
$\mathbf{120}$  FM Higgs representation(Next to Minimal Susy GUT
or NMSGUT). The $\mathbf{120}$ had been  optimistically excluded
in the Babu-Mohapatra program to foster the predictivity of the
MSGUT in the neutrino sector. In this alternative
scenario\cite{blmdoom} the $\mathbf{120}$ -plet collaborates with
the $ \mathbf{10}$ -plet to fit the dominant charged fermion
masses. The small $\mathbf{\oot} $ -plet couplings give
appreciable contributions only to light charged fermion masses
{\textit{and}} enhance the Type I seesaw masses  to viable values
since the Type one seesaw masses are \emph{inversely} proportional
to these masses. To begin analysis of the fitting problem in this
new and qualitatively different fitting scenario we
analyzed\cite{nmsgutI}  the 2-3 generation case as  a toy model
 of   the  dominant core of the
complete hierarchical fermion mass system. We found that
\emph{ansatz} consistency \emph{requires}    $ {\mathbf{m_b-m_s =
m_{\tau}-m_{\mu}}} $  at the GUT scale $M_X$ and predicts near
maximal (PMNS) mixing in the leptonic sector for central values of
charged fermion parameters and for   wide ranges of the other
relevant parameters (righthanded neutrino masses and relative
strength of contributions of the two doublet pairs from the
$\mathbf{120}$-plet to  the effective  MSSM Higgs pair).

In the current contribution  we report on progress in extending
the analysis of \cite{nmsgutI} to a 3 generation CP conserving but
otherwise   realistic NMSGUT model. We first briefly review the
failure of the seesaw mechanism in the MSGUT and the  results of
our analysis of the 2-3 generation toy model.  This is followed by
a  description of the fitting equations in the CP conserving 3
generation case and our procedure for obtaining accurate solutions
of the 3 generation equations by an effectively  linear
perturbation expansion (in the Wolfenstein parameter $\epsilon
\sim \sin\theta_{12}^c\sim\sqrt{\sin \theta_{23}^c}\sim .2$)
around the dominant  2-3 generation  mass matrix structures. We
then present examples of solutions that are realistic in all
respects except that of the omitted CP violation and the lack of
two large mixing angles. We argue that since even in the
$\mathbf{10-\oot}$ Type I seesaw a solution with two large angles
and the observed small ratio of mass splittings\cite{strumviss}
could be found only when CP violation was also
included\cite{matsu0,babmacesanu} our results should be construed
positively. The study of the CP violating case is now in
progress\cite{nmsgutII}. Thus the main obstruction of too small
Type I masses is already overcome even in the CP conserving model
but   the fit for the CP violating case remains to be done.

\section{  Seesaw Failure   in the MSGUT}

  The Type I and Type II
 seesaw  Majorana masses of  the light neutrinos  in the MSGUT are\cite{blmdoom}  :

\bea M_{\nu}^I &=& (1.70 \times 10^{-3} eV) ~ { F_{I}}~
\hat{n}~{\sin \beta}\nnu
 M_{\nu}^{II} &=& (1.70 \times 10^{-3} eV) ~{ F_{II}}
 ~\hat{f}~{\sin \beta}\nnu
    \hat{n}&=& ({\hat h} -3 {\hat f}) {\hat f}^{-1}
    (  {\hat h} -3 {\hat f})\nonumber
     \eea

where $v=174~ GeV$  and  $\hat{h},\hat{f}$ are
 the Yukawa coupling matrices of $\mathbf{10,\oot}$ to the
$\mathbf{16}$ plets containing fermion families and  $\beta$ is
 the MSSM Higgs doublet
mixing angle. The functions $F_I,F_{II}$ are completely specified
by in the MSGUT and explicit forms may be found in
\cite{ag2,gmblm,blmdoom,bmsv2}.

In typical BM-Type II fits\cite{gohmoh,bert,babmacesanu} the
maximal value of $ \hat{f}$ eigenvalues is $ \sim 10^{-2}$ while
the corresponding values for $ \hat{h} $ are about $10^2$ times
larger.  As a result  $\hat{n}_{max} \sim 10^2  $.
 This implies that  $R={F_I/F_{II}}  \leq 10^{-3}$ in order that
  the pure BM-Type II    not be overwhelmed by
   the BM-Type I values it implies. Such R values were
    shown\cite{gmblm,blmdoom} to be un-achievable anywhere
    in parameter space of the MSGUT (while preserving
  baryon stability, perturbativity etc).    Furthermore
in the BM-Type I fit of \cite{matsu0,babmacesanu} the requirement
of large neutrino mixing yields, typically, $ \hat{n}_{max} \sim 5
\hat{f}_{max} \sim .5  $ for the maximal eigenvalues of $
\hat{n}$. Thus values of $F_I\sim 100$ are required in order to
reach realistic values of seesaw masses for the heaviest neutrino.
We  demonstrated\cite{gmblm,blmdoom} that such values are not
achievable anywhere over the  complex $x $ plane except where they
also violate some aspect of successful unification. Recently
another group has verified our results\cite{bertnu}

\section{The new $\mathbf{10-120-\oot}$ scenario }

    To resolve the difficulty with the overall neutrino mass sscale
 we proposed\cite{blmdoom,nmsgutI}
    that the $\mathbf{\oot}$ couplings be reduced
    far below the level where they are important for 2-3 generation
    masses  while   introducing a $\mathbf{120}$ plet to do the work of
    charged fermion mass fitting previously accomplished by ${\bf{\oot}}$
  couplings  almost comparable to those of the ${\bf{10}}$-plet.

  The Dirac masses in such GUTs are then
generically given by\cite{ag1,ag2,blmdoom,nmsgutI}
\bea \hat m^u &=&  v( {\hat h} + {\hat f} + {\hat g} )\nnu
 \hat m_{\nu}&=&v ({\hat h} -3 {\hat f}  + (r_5 -3) {\hat{g}})\equiv
 v ({\hat h} -3 {\hat f}  + r_5' {\hat{g}})
 \nnu
\hat m^d &=& { v (r_1} {\hat h} + { r_2} {\hat f}  +
r_6 {\hat g}) \\
   \hat m^l &=&{ v( r_1} {\hat h} - 3 {  r_2} {\hat f} + r_7{\hat g})
       \label{120mdir}\nonumber\eea

  See \cite{ag1,ag2,blmdoom,nmsgutI} for the form of the   coefficients
$r_i$.

 The right handed neutrino mass is $M_{\bn}  = \hat{f}
\hat{{\bar{\sigma}}}$ and the Type I seesaw formula is \bea
M_{\nu}^{I} =   v r_4 \hat{n} \quad ;\quad \hat{n}&=& ({\hat h} -3
\hat{f} - r_5' {\hat g}) {\hat f}^{-1}
    (  {\hat h} -3 \hat{f }+ r_5'{\hat g})\eea
 where $\hat{{\bar{\sigma}}}     = \frac{i\bar{\sigma}\sqrt{3}}
    {\alpha_2 \sin \beta} $, $\bar{\sigma} $
is the GUT scale  vev of the $\oot$ while  $\alpha_i,\bar\alpha_i
$ refer to fractions of the MSSM doublets contributed by various
doublets present in the GUT Higgs representations
\cite{abmsv,bmsv,ag2,gmblm,blmdoom}.

The essence of our proposal is   that the  ${\bf{10,120}}$
multiplet Yukawas dominate the contributions of the $\bf{\oot}$
coupling in the charged fermion masses i.e
$(\hat{h},\hat{g})_{max}>>\hat{f}$. We shall work with the
assummption that the fermion  Yukawa couplings and coefficients
$r_i$ are both real. The latter assumption is almost certainly not
valid\cite{ag1,ag2,bmsv}. However if a fit could be found with all
complexity introduced via these coeffcients and the Yukawa
couplings were kept real then the resulting NMSGUT would contain
only 12 Yukawa coupling parameters
 i.e 3 less than in the MSGUT with complex Yukawas. Thus even with
 the new couplings due to the $\mathbf{120}$  the NMSGUT with only spontaneous
  CP violation would still be a strong contender as far as minimality
   of parameters is considered as decisive.

The  mass terms above  must be  matched to the
 renormalized mass matrices of the
MSSM evaluated at the GUT scale. While doing so one must
 allow\cite{gmblm}
 for the  possibility  that the fields of the GUT are only unitarily
 related to those  of the MSSM at $M_X$. This introduces several
   unitary matrices   into the fitting problem
   which, besides conventional ambiguities, specify just
    how the MSSM lies within the MSGUT once conventions are fixed. Although
   unphysical in the MSSM these matrices are of vital relevance
 for calculation of the exotic signatures (e.g B-L violation) of
 the GUT in question\cite{gmblm}.
 Thus when matching the  charged fermion dirac mass matrices   we get

\bea {\hat m}^u &=&
  V_u^T  {D}_u Q\nnu
 {\hat m}^d &=&  V_d^T  {D}_d R \nnu
{\hat m}^l &=&  V_l^T  {D}_l L
   \eea

Where $D_{u,d,l}$ are the charged fermion masses at $M_X$  and
$V_u,Q,V_d,R=C^{\dagger} Q,L,V_l$ are arbitrary unitary matrices
($C$ is the CKM matrix). These matrices must  be fixed by
convention --where allowed by conventional ambiguities -- or
determined by the fitting procedure in terms of the low energy
data, or left as parameters to be determined by future experiments
sensitive to degrees of freedom and couplings (e.g baryon
violating couplings) that the low energy data is not.  We shall
make the SO(10) basis choice $R=1$ in what follows.

To put our equations in a form transparent enough to clearly
separate the contributions of the $\mathbf{10,120 }$ (and
-eventually-  $\mathbf{\oot}$) plets\cite{nmsgutII} we write the
matrices $V_{u,d,l}$ associated with the anti-fermion fields  as
new unitary matrices $\Phi_{u,d,l}$ times the fermion field
matrices $Q,R,L$. \bea V_d &=& \Phi_d R \hspace{5mm}; \hspace{5mm}
V_u = \Phi_u  Q \hspace{5mm}; \hspace{5mm} V_l = \Phi_l L \eea We
then separate symmetric and antisymmetric parts for each charged
fermion equation
\bea Z &=& \Phi^T {D} + {D} \Phi \hspace{5mm};
\hspace{5mm} A = \Phi^T D - D \Phi  \eea

Solving the mass formulae for the symmetric yukawa couplings one
finds :

\bea \hat h &=& {\frac{\hat r_1 }{2v }} R^T (3 Z_d + {\cal D} Z_l
{\cal D}^T ) R \nnu  \hat f &=& {\frac{\hat r_2 }{2v }} R^T (  Z_d
 - {\cal D} Z_l {\cal D}^T ) R  \eea

while the antisymmetric couplings are

\be \hat g={\frac{1}{2 v r_7}} R^T  A_l R\ee

and where $\hat r_i= 1/(4r_i)$.

Then the remaining equations are the ``sum rule''

\bea (3 {\hat r_1 } + {\hat r_2 } ) Z_d  + ( {\hat r_1 }  - {\hat
r_2 } )   {\cal D} Z_l {\cal D}^T  - C^T Z_u C =0 \label{sumrule}
\eea

\subsection{23 generation toy model}

We first briefly recapitulate the analysis of the 2 generation CP
conserving toy model given in \cite{nmsgutI}. As explained, in
this case the contribution of $\hat f$ to the charged fermion
masses is taken to be negligible. Then  using (say) just the
d-type quark equations to solve for  $\hat h,\hat g$ one obtains
for the rest :

 \bea C^* Z_d{C}^{ \dagger}&=& r_1 Z_u\hspace{5mm} ;
 \hspace{5mm} Z_d = {\cal D} Z_l {\cal D}^T \hspace{5mm};\hspace{5mm}
{\cal D} = R^* L^T  \nnu C^* A_d{C}^{ \dagger}&=& r_6 A_u \hspace{5mm} ;
\hspace{5mm}r_7 A_d = r_6 {\cal D} A_l {\cal D}^T \label{aseq}\eea

In this 2 generation case the  antisymmetric
equations(\ref{aseq}) serve only to fix the parameters $r_6,r_7$
and thus play no further role. On the other hand since $C,{\cal D}$  are
orthogonal matrices it follows that $r_1=Tr(Z_d)/Tr(Z_u)$   so
that we can write these two equations
 in the (dimensionless)  form

\bea
\widehat{S}_1 &=& \frac{C Z_d C^T}{Tr Z_d}  - \frac{  Z_u  }{Tr Z_u}=0\nonumber\\
\widehat{S}_2 &=& \frac{Z_d - {\cal D} Z_l {\cal D}^T}{Tr Z_d}=0
\eea

  In \cite{nmsgutI} we gave an analytic solution of the toy model
 charged fermion mass fitting relations.
We parametrized the matrices  $\Phi_{u,d,l},C, {\cal{D}}$ as \bea
\Phi_{u,d,l} &=& \left( \begin{array}{cc}\cos\chi_{u,d,l}
\hspace{5mm}\sin\chi_{u,d,l}\\
-\sin\chi_{u,d,l} \hspace{5mm}\cos\chi_{u,d,l} \end{array} \right) \nonumber\\
 {C,\cal{ D}} &=& \left( \begin{array}{cc}\cos\chi_{c,\cal{D}}
\hspace{5mm}\sin\chi_{c,\cal{D}}\\
-\sin\chi_{c,\cal{D}} \hspace{5mm}\cos\chi_{c,\cal{D}} \end{array}
\right)  \eea

 Then we found that the equations $\hat S_1=0$ yield ($\alpha  ={\frac {(d_3-d_2 )
(u_3+u_2)}{ (d_3+d_2) (d_3-u_2)}}$)

\bea \chi_u &=&   \chi_d - 2 \chi_c\nnu \tan \chi_d &=& \frac{\csc
2 \chi_c}{\alpha}- \cot 2\chi_c \label{chid}\eea

  Since $\alpha=1+(\epsilon^2)$, it is
clear that the leading ($\sim \epsilon^{-2} $) contributions
cancel leaving behind an $O(1)$ result.
 Similarly two of the  \emph{three}  equations $\hat{S}_2=0$ yield

 \bea\chi_D&=&(\chi_l-\chi_d)/2 \qquad;\qquad
\chi_l=\pm\bar{\chi}_l \nnu \bar{\chi}_l&\equiv&\cos^{-1} {\frac{T_d
\cos \chi_d}{T_l}}\eea
To leading order $\chi_l =\pm \chi_d$, this defines two branches
of the solution which we call the (+) and (-) branches following
the sign between $\chi_l$ and $\bar\chi_l$. The third $\hat S_2=0$
equation yields the all important consistency condition
 \bea { {d_3 - d_2  =\Delta d = \Delta l  = l_3 -l_2 }} \eea
i.e
\bea   {{m_b(M_X)  - m_{\tau}(M_X)  = m_s(M_X)  - m_{\mu}(M_X)
}}\eea

Using this solution we determined the 23 sector  PMNS mixing angle
as a function of the parameters $r_5',\rho=\hat f_3 /\hat f_2$
($\hat f_i$ are the eigenvalues of $\hat f$) and then showed that
the 23 sector PMNS mixing  was near maximal over much of the parameter
space of the theory. Thus the toy model provides a
cartoon of the core of the 23 generation dominated fermion
hierarchy that is satisfactory and coherent with our scenario in
all respects.

\subsection{3 generation fitting formulae}

We now wish to generalize our 2 generation toy model to the 3
generation case, while keeping the CP conserving approximation for
simplicity, to enquire how closely one may approach the actual
fermion data. The resulting equations are however too complicated
to solve analytically. Thus one must resort to some variety of
numerical approximation technique or perturbation theory. Our
approach is to consider the 23 sector as the dominant ``core'' of
the fermion mass hierarchy and thus to expand around this core by
rewriting all the charged fermion mass parameters and mixing
angles relative to the dominant normative magnitudes of the third
generation  by introducing a small parameter $\epsilon \sim
{\sqrt\chi_c}\sim \sin{\theta_{12}^c}\sim
\sqrt[3]\theta^c_{13}\sim \sqrt {d_2/d_3} $ etc.  The explicit
analytic solution  of the toy model   given above coincides, order
by order in perturbation theory , \emph{including the
$\mathbf{b-\tau=s-\mu}$ constraint }, with the solution found
   by expanding all angles $\chi$ in powers of
$\epsilon$  and solving the equations $\hat{S}_1=0=\hat{S}_2 $
 order by order in $\epsilon $. The  $\mathbf{b-\tau=s-\mu}$
 unification constraint arises at order  $\epsilon^2$.
  In the three generation case the analytic solution is not available but
 the complete regularity observed in the perturbation theory in
 $\epsilon $  and the extreme smallness of the first
generation perturbations  makes an expansion in $\epsilon$ well
motivated. Thus we shall cast the fitting equations in a form such
that they reduce to the toy model equations up to order
$\epsilon^2$ since $\hat f, d_1,u_1,l_1$ appear only at order
$\epsilon^3$ or even higher.

 We continue with our CP conserving restriction to maintain
 tractability and parametrize each of the  orthogonal
matrices   in the standard Kobayashi-Masakawa form for real
unitary, i.e orthogonal, matrices :

\bea O  &=&  O_{23}(\chi) \cdot O_{13}(\phi) \cdot O_{12}(\theta)
\eea

where $\chi,\phi,\theta$ are generic symbols for rotation angles
in the $23,13,12$ sectors. We shall always work in the SO(10)
basis  fixed by the condition  $R=1$.

Now the antisymmetric equations are no longer empty but read

\bea  {\hat A}_{u,d,l}    &\equiv&   {\frac{A_{u,d,l}} {\sqrt{{\vec
A_{u,d,l}} ^2} }} \nnu \hat{A_1} &=&  C  \hat A_d  C^T   - {\hat
A}_u =0\nnu
 {\hat A}_2^{\pm} &=& \hat A_d  ~ \mp ~ {\cal  D } {\hat A}_l   {\cal D
 }^T=0\label{ahats}\eea

where

\be  {\vec A}^2= A_{12}^2  + A_{13}^2 + A_{23}^2\ee

and the sign in the last of  equations(\ref{ahats})  is chosen
corresponding to the branch of the 2 generation model that one is
expanding about i.e   $\hat{A_2}^{\pm}=0$   for $\chi_l=\pm
\bar\chi_l$. This is because   we  solved for the coefficients
$r_6^2,r_7^2$ in terms of square roots of ratios of $Tr A_f^2$
($f=u,d,l$). So we should retain an option for the sign of the
square root : which must be exercised as noted, for consistency,
when considering the equations appropriate for expansion around
the two branches of the toy model solution discussed above. The
necessity of our choice can be  easily understood by considering
the   two generation example where it can be
   easily checked that ${\hat A}_2^{+}$ cannot vanish on the (-) branch.

Defining

\bea \hat X &=& {\frac{3 Z_d +{\cal D} Z_l {\cal D}^T }{3 Tr Z_d +
Tr Z_l}}\nnu \hat S_X &=& \hat X - C^T {\frac {Z_u}{Tr Z_u}} C
\eea

we have finally

 \bea
 \hat S_3 & = & \hat S_X
     + {\hat r_2}  {\frac{Tr Z_d}{Tr Z_u}}
 ( {\hat S_2}  - {\frac{(Tr Z_d - Tr Z_l)}{ Tr Z_d}}{\hat X})  \eea

Since the up sector fermion masses are about $10^2$ times  larger
than the down or charged lepton sector it is convenient to rescale
the values of the up sector masses  by   $10^2$ : $Tr Z_u=10^2
\tilde Tr Z_{\tilde u}$ and $\hat r_2 = 10^2 \tilde r_2$. The
definitions are chosen so that when $\hat S_2=0$ then $\hat S_3=0
$  reduces to $\hat S_1=0$. Then, in view of the encouraging
results of our 2 generation model for the 23 dominated fermion
hierarchy,  it is natural to look   for solutions to the fitting
problem not \emph{ab initio}, i.e not for a solution of hopelessly
non-linear equations but by an expansion (in powers of
$\epsilon\sim \sqrt{\theta_{23}}\sim \theta_{12}\sim .2 $) around
the 2 generation results. Since the 2 generation case gave $\hat
S_2\sim O(\epsilon^3)$ it follows that   :

 \be \hat f =  {\frac{1}{2v}} {\hat r_2} Tr Z_d {\hat S_2}
 \ee

 The $O(10^{-2})$ suppression provided by the ratio
$d_3/v$ in the above equation   implies   that $(\hat f)\sim .01\epsilon^{3+\delta}$ when
   ${\hat r_2}\sim \epsilon^{\delta}$. This  will then ensure the
enhancement of Type I neutrino masses that is the rationale for
this fitting scenario.

 We chose the SO(10) basis where $R=1$ and  expand all the
 orthogonal matrix angles  $\theta_{u,d,l,{\cal D}},\phi_{u,d,l,{\cal D}},\chi_{u,d,l,{\cal
 D}}$ as well as the free parameter $\tilde r_2$ in powers of  $\epsilon
 $. The expansion is started by assuming either

\be \chi_d^{(0)}=\chi_u^{(0)}=\chi_l^{(0)}=\chi_{\cal D}^{(0)} \ee
  corresponding to the $\chi_l=+\bar\chi_l $
 solution,   \emph{or }

 \be \chi_d^{(0)}=\chi_u^{(0)}=-\chi_l^{(0)}=-\chi_{\cal
  D}^{(0)}\ee

    corresponding to the $\chi_l=-\bar\chi_l $ solution.
    Furthermore, since the off-diagonality of the sum rule eqn.(\ref{sumrule}) is
    driven by that of the CKM matrix which has $\phi_c\sim \epsilon^3$
    we make the weak and well justified assumption (further justified post hoc by
    actually finding solutions of this type)
     that the angles $\phi_{u,d,l,{\cal D}}$ are $O(\epsilon)$ or
     smaller and thus their expansions begin at order $\epsilon$.

For order $\epsilon^{0,1,2}$ the expansion simply reproduces the
expansion of the toy model already described above in Section 3
together with some additional constraints on the new angles. For
example for the $(+)$ type solutions, at order $\epsilon^0$, one
gets $\theta_u^0=\theta_d^0, \theta_{\cal
D}^0=\theta_l^0-\theta_d^0$. At every order of perturbation  we
always solve for parameters in which the equations at that order
are \emph{ linear} and of as high order as possible. This obviates
the difficulty of multiple solutions of nonlinear equations and
their extreme sensitivity to the grossly mismatched coefficients
implied by the fermion mass hierarchy which stretches over 5 order
of magnitude. At every order beyond $\epsilon^7$ (where the
smallest of the mass parameters i.e $\tilde u_1\sim \epsilon^7$
enters the $\hat S_3=0$ equation) we evaluate the $\chi^2$
parameters for the charged fermion fit by setting all undetermined
parameters to zero. The expansion is pursued till as high as 20th
order (!) and the solutions with smallest $\chi^2$ retained. Then
using the truncated values at the chosen order and the values of
the determined expansion coefficients we reconstruct the fermion
mass data via the reconstructed Yukawa couplings  and mass
relation coefficients.   We defer a report of the details of these
solutions to \cite{nmsgutII} but two points are crucial : the
value of $\chi_d^0$ and the constraint $d_3= l_3-l_2 + d_2 $ are
the same as in the toy model. Thus in line with our discussion of
that model we shall fit the central values of the charged fermion
masses and mixing angles \emph{except for the value of $d_3$ which
we shall take to be exactly} $d_3=m_{\tau}-m_{\mu} + m_s $.

Till order $\epsilon^2$ the equations are $\hat S_1=\hat S_2 =0$,
thereafter since $\tilde r_2 \sim \epsilon^{3 +\delta}$ if $\hat
r_2 \sim\epsilon^{\delta}$ it follows that the complete equation
$\hat S_3=0$ reduces to $\hat S_X=0$ till one reaches order
$\epsilon^{6 +\delta}$. Thereafter the full equation  $S_3=0$ must
be used with the expansion of $\tilde r_2$ beginning at order
$\epsilon^{3+\delta}$. Since the up quark mass enters the
equations only at order $\epsilon^7$ there is no question of
finding accurate solutions before that order. Furthermore many of
the angle expansion coefficients are left undetermined at any
given order after all the equations available at that order have
been solved. These undetermined coefficients are simply set to
zero before evaluating the results numerically. Note that this
cannot affect the Orthogonality of the matrices in which the
relevant angles occur and that this luxury is a consequence of our
parametrization of the sum rule. Such a procedure may look
mathematically questionable particularly since the convergence
appears to be very weak and typically the values of the masses and
mixing angles \emph{oscillate } in an apparently haphazard manner.
The reason for this may actually be the absence of  complexity or
the poorness of $\epsilon$ as an analytic expansion parameter.
Nevertheless the proof of the pudding lies in the eating of it !
Our aim -as that of any competing numerical fitting procedure - is
to find a ``close'' fit i.e to obtain values of the coefficient
matrices $\hat f,\hat g,\hat h$  and coefficient quantities
$r_1,r_2,r_5,r_6,r_7$  such that when we use the generic form of
the  SO(10) GUT mass formulae we generate charged and neutral
fermion masses compatible with the low energy fermion data
extrapolated over an MSSM desert to the GUT scale. As in the
fitting done using the downhill simplex method
\cite{grimuslavu,bertnu} a simple criterion to pick out an
acceptable solution   is  to define    $\chi^2 $ functions and ask
that they be appropriately small :

\bea \chi^2_{m} &=& \sum_i ({\frac{m_i-\bar m_i}{\delta m_i}})^2
\nnu \chi^2_{CKM} &=& \sum_a
({\frac{\sin\theta_a-\sin\bar\theta_a}{\delta\sin\theta_a}})^2\\
\chi^2_{tot}&=&\chi^2_{m} +\chi^2_{CKM}\nonumber\eea

For the central quark masses and angles we used the Das-Parida
central values the fermion data (for $\tan\beta(M_S) =55$ and at
$M_X=2\times 10^{16} GeV$ ) (except for $d_3$ as explained above)
:

\bea {\bar m}_u  &=&  .000724 \quad ; \quad {\bar m}_c   = .2105
\quad ; \quad {\bar m}_t = 95.148\nnu
 {\bar m}_d  &=&  .001497\quad ; \quad
{\bar m}_s  =   .0298 \quad ;
\quad {\bar m}_b = {\bar m}_\tau + {\bar m}_s - {\bar m}_\mu=1.5835    \\
 {\bar m}_e  &=& .000356 \quad ; \quad
{\bar m}_{\mu}   =   .0753 \quad ; \quad {\bar m}_{\tau}  =
1.629\nnu
 \sin \theta_c & =&
.2248 \quad ;\quad
 \sin\chi_c = .03278 \quad ;\quad
\sin\phi_c = .00216 \nonumber \eea

For the $1-\sigma$ error estimates we used (all masses in GeV) :

\bea    \delta  m_t  &=&  40.0 \quad ; \quad  \delta  m_c  =   0.018
\quad ; \quad \delta  m_u = .00013 \nnu
  \delta m_b &=& 0.34 \quad ; \quad \delta m_s = .0042\quad ; \quad  \delta m_d = .0003 \nnu
   \delta m_\tau &=&
.038 \quad ; \quad \delta m_\mu = 0.00013 \quad ; \quad \delta m_e
= 0.0000015\nnu \delta
 \sin \theta_c &=& .0016 \quad ; \quad \delta \sin\phi_c  = .0005 \quad ; \quad
  \delta  \sin\chi_c  = .0013 \eea

For each parameter at an average of (say) half a $1-\sigma$
deviation away from the central value one might expect a
contribution to $\chi^2$ of , roughly $.25$ so that
$\chi^2_{m}\sim 2.25 , \chi^2_{CKM}\sim.75$. Determining fits much
more accurate than these values is academic as regards the actual
values. It is informative only inasmuch as the ability to find
finely matched fits to an \emph{a priori} un-sacrosanct  set  of
central values for only approximately known parameters implies
only that  \emph{any} such set of data is likely to be achievable
via a fit of the type indicated. In view of the fact that we
pushed $m_b$ up from its central value by about $\delta m_b/2$ to
begin with one might even expect larger deviations from the
``correct'' values (the quotes reflect the fact that the internal
correlations of central values and errors in the extrapolated data
at $M_X$ are at present very poorly understood). Moreover since
the correct scenario for 3 generations must also explain CP
violation, it may be that the slow and oscillatory convergence we
observe is due to an obstruction in the approach to the true
solution due
 to a projection onto a real subspace of the actual parameter space.   Surprisingly our weakly convergent
perturbative method frequently gives solutions with $\chi^2_{tot}$
as small as $.2 $ while  solutions with  $\chi^2\sim 2$ are hardly
difficult to find. In the next section we discuss examples of the
solutions found and of how much closer they have brought us to
realizing our scenario of a fully successful, NMSGUT compatible,
fermion data fit.

\subsection{Neutrino Masses and Mixing}

 As shown and discussed in \cite{gmblm,blmdoom,nmsgutI} the  the Type I seesaw mass formula
 may be written
\bea \hat M_{\nu}^I&\simeq & \frac{v^2}{2\widehat{\bar{\sigma}} }
(\widehat{h}+ r_5^{'}\widehat{g}-3 \hat f)^T
\hat{f}^{-1}(\widehat{h}+ r_5{'}
\widehat{g} -3 \hat f)\nonumber\\
  &\equiv&(1.70 \times 10^{-3} eV) R^T \hat n R F_I \sin \beta
\nonumber\\
&=&  {L}^T \mathcal{P} D_{\nu}\mathcal{P}^T  {L} \label{TypeI}\eea
Where ${\cal{P}}$ is the Lepton mixing (PMNS)
matrix\cite{pontemaki} in the basis with diagonal leptonic charged
current and  $D_{\nu}$ the light neutrino masses extrapolated to
$M_X$.

The PMNS matrix ${\cal P}$ and the ratio of solar to atmospheric
mass squared splittings can then be identified as
  \bea {\cal{P}}& =& D^{\dagger}
{\cal{N}} \qquad ;\qquad
{\frac{{m}_{sol}^2}{{m}_{atm}^2}}={\big|}{\frac{\hat{n}_{\mu}^2
-\hat{n}_e^2}{\hat{n}_{\mu}^2 -\hat{n}_\tau^2}}{\big|}
\label{DTF}\eea

where ${\cal{N}}$ diagonalizes $\hat n : \hat n= {\cal{N}}\hat
n_{diag} {\cal{N}}^T$. Since it is very reasonable to assume that
the coefficient function $F_I$ will generically have the same
magnitudes in the NMSGUT as it did in MSGUT i.e $\leq 1$, it
follows that we can can verify whether the enhancement of $\hat n$
makes the NMSGUT generically viable simply by comparing it with
the typical values $\hat n_{max}\sim .5$ which failed (for generic
$F_I$ values) in the MSGUT case by about two orders of magnitude
and for special ones by about one order of magnitude.

\section{Examples of  3 generation  fits}

Our purpose in this contribution is no more than to outline the
likely features of 3 generation fits that follow the scenario
suggested\cite{blmdoom,nmsgutI} by us as a route to overcoming the
generic debility of neutrino masses \cite{gmblm,blmdoom}  in the
Babu-Mohapatra\cite{babmoh} program.
 Therefore we here  only  give as examples  a few  quasi-realistic fits, of the type
sought, and a few others for comparison, obtained  using the
method outlined in the previous section. As described there we
simply took the central values of\cite{dasparida} and, after
implementing the constraint of $b=\tau+s-\mu$ unification
\cite{nmsgutI} required by the consistency of the fitting
equations,  aimed for the central values of the other parameters.
Thereafter we fit to the neutrino data as closely as possible.
This is what we report here. No attempt was made by us as yet to
search the parameter space for ``optimal'' fits. A  detailed
numerical survey will be reported
 in\cite{nmsgutII}. Of course improvement of the
procedure regarding the connection to the low energy data is also
possible, but, in our opinion this will be called for  only after
the gross features of the viable fermion data fit are established
and CP violation is included. In our view this fit should not be
and is not so unstable and delicately poised as to change its
qualitative features due to shifts (due to better RG flow control)
of a few per cent in the central values that one is trying to fit.

  Our procedure was first to fit the ratio of neutrino mass
squared splittings to the current\cite{strumviss} central value of
the solar to atmospheric mass squared splitting ratio $\Delta
m^2_{sol}/\Delta m^2_{atm}\sim .032$ by adjusting and thus fixing
the value of $r_5'$. This gave a number of candidate $r_5'$ values
for a given charged fermion data fit. Each value $r_5'$
corresponds to a certain neutrino mass hierarchy and determines a
certain value for the eigenvalues of $\hat n$ : the crucial matrix
whose maximum eigenvalue and coefficient in the Type I seesaw mass
were generically found\cite{gmblm,blmdoom} to be too small in the
MSGUT. Using the values of the angles of the orthogonal matrix
${\cal D}$ found as part of the solution procedure it is easy to
generalize the formulae of Section 3 and obtain the three PMNS
mixing angles.

\subsection{$(+)$ Type fit}

In the case of the $(+)$ type solutions we find that the solutions
obtained assuming $\hat r_2\sim \epsilon^{0,1,2}$ all yield final
$\hat r_2 $ values that  differ between the solutions with lowest
$\chi^2$ by  no more than about $15\%$. The $\chi^2$ values are
also very similar ( $\chi^2_{tot}\sim .14$ or about .01 per
variable ! : an excellent fit by normal standards). Since the
effect of $\hat r_2$ on the other quantities in the fit is
\emph{extremely} weak, the resultant values of the Yukawa
couplings and  fermion masses and angles are almost identical.
This strengthens our notion of a unique stable realistic
solution/map, of the experimentally found hierarchical type, tying
together the Yukawa coupling matrices and coefficients to the
fermion data. Nevertheless a detailed survey is required to
determine the actual ranges of mixing angles and neutrino
mass-squared splitting obtainable and even perhaps alternative
branches of fits that may actually  arise.   The values given
below have been truncated to 4 decimal places but the actual
calculation was done retaining 10 digit accuracy. In the
representative $I+$ case where $\hat r_2$ is assumed to be $\sim
\epsilon $ to begin with we find (${\hat r}_i=(4 r_i)^{-1}$)

 \bea {\hat r}_1 &=& 15.27  \quad ; \quad
            {\hat r}_2 =0.255 \quad ; \quad r_6=0.0187 \quad ; \quad
          r_7=0.023305 \nnu
          \nnu
           {\hat{h}}  &=&
\left(\begin{array}{ccc}
 -0.0002988&0.0000394&-0.0222691\\ 0.0000394&
    0.009291&-0.134823\\-0.0222691&-0.134823&0.481333
  \end{array} \right)\nnu
 \nnu{\hat{f}}  &=&
\left(\begin{array}{ccc}0.0000132&0.0000277&-0.0001\\0.0000277&-0.0000146&
    0.000057\\-0.0001&0.000057&0.000040
  \end{array} \right) \\
\nnu{\hat{g}}  &=& \left(\begin{array}{ccc}0& 0.001644& -0.02478 \\
 -0.00164 &
      0.& -0.119710\\0.02478&0.119710&0
  \end{array} \right)\eea

The eigenvalues of $\hat h,\hat f,\hat g$ are

\bea \hat h &:& .518~~\quad;~~.0277~~\quad;~~1.92\times 10^{-5}
\nnu \hat f &:& 1.33\times 10^{-4} ~~;~~1.11\times
10^{-4}~~;~~0.171\times 10^{-4}\nnu \hat g&:& \pm 0.122 ~~;~~0\eea

Note how the premises of  our scenario\cite{blmdoom,nmsgutI}  are
indeed respected.

  Then the
reconstructed values of the charged fermion masses are(in GeV)

 \bea M_U &=& \{95.148,0.2105,0.00077\}\nnu
 M_D &=&\{1.5835,0.02981,0.0015\}\nnu
M_l&=& \{1.629,0.0753,0.000356\}\eea

note that we always fit the charged lepton masses exactly. The
reconstructed CKM matrix is
 \bea {\hat{C}}  &=&
\left(\begin{array}{ccc}
0.974449&-0.224599&0.002155\\0.224548&0.973911&-0.032784\\0.00526452&
    0.0324303&0.99946
  \end{array} \right)\eea

 Which yields the  CKM angle magnitudes

 \be \theta_{12} =0.2265 \quad ; \quad \theta_{13} =0.002155\quad ;
  \quad \theta_{23} =0.03279 \ee

Then one finds for this solution

 \be \chi^2_m= 0.126  \quad ; \quad \chi^2_{CKM}=.016 \quad;
 \quad \chi^2_{tot}= 0.142   \ee

To determine the neutrino mixing data the free parameter $r_5'$
must be fixed. If it is known one can calculate $\hat n $ and then
since ${\cal D}$ is known from the solution the mixing angles are
easily determined.  We  fix $r_5'$    by enquiring for which
values of $r_5'$ the six possible ratios

\be R_{ijik}= \big|{\frac{{\hat n}_i^2 -{\hat n}_j^2}{{\hat
n}_i^2-{\hat n}_k^2}}| \ee

can attain the current\cite{strumviss} central value
$R=m^2_{sol}/m^2_{atm} =.32$. It is then evident that then the
indices  j,i,k  correspond to the $e,\mu,\tau$ neutrinos
respectively and so we labelled such a solution as an $ijik$ type
solution. The software we used allots the index 1 to the largest
eigenvalue, 2 the next largest and 3 the smallest, this accounts
for the solution labels encountered below.

 For the I+case in question one
finds the following set of solutions :

\bea I+{2321}   &:&~(a)~r_5' = -0.33285 \quad ;\quad ~ (b)~ r_5'
=0.48265\nnu I+{3231}  &:& ~(a)~ r_5'= -0.33140 \quad ;\quad ~
(b)~ r_5'=0.481314\nnu I+{2123} &:&~ (a)~r_5'=-0.14427 \quad
;\quad ~(b)~r_5'=-0.13932\nnu &&~(c)~r_5'= 0.301785 \quad ;\quad
(d)~ r_5'=0.30650\nnu I+{1213} &:& ~(a)~ r_5'=-0.14435 \quad
;\quad(b)~ r_5'= -0.139237\nnu && ~(c)~r_5'= 0.306578 \quad
;\quad~(d)~r_5'=0.30171\eea

It is clear that the cases I+2321 and I+3231 and cases I+2123 and
I+1213 are almost identical.   Note also how close  I+2123(a) and
I+2123(b)   and I+2123(c)  and I2123(d) are. In fact these pairs
give almost identical leptonic mixing parameters and $\hat n$
eigenvalues.Case I+2321 corresponds to the hierarchy $m_{\nu_\tau}
>> m_{\nu_\mu}>> m_{\nu_e}$ while case I+3231 corresponds to
$m_{\nu_\tau} >> m_{\nu_e} >> m_{\nu_\mu}$. Similarly case I+2123
corresponds to the hierarchy $m_{\nu_e}
>>m_{\nu_\mu}>> m_{\nu_\tau} $ and case I+1213 to $m_{\nu_\mu}>>m_{\nu_e}
>> m_{\nu_\tau} $. Cases I+2321,I+3231 have highly hierarchical $\hat
n$ eigenvalues while in cases  I+2123,I+1213 which have an
inverted hierarchy there are two large almost degenerate
eigenvalues and one very small one. Because of the close
similarities   we give only the values for cases I+2321a,I+2321b
and I+2123a,I+2123c in Table 2. Only in case I+2312 is the $23 $
sector mixing angle large and the $13 $ mixing angle reasonably
small. \emph{However the value of the 12 sector mixing is very
low}. Clearly then two large and one small PMNS mixing angles are
hard to achieve. On the other hand the large ( about 200 times
larger than the Type I fits in the $10-\oot$ scenario) value of
the largest $\hat n$ eigenvalue together with the satisfactory
value of the mass squared splitting ratio means that the problem
with too small neutrino masses is unlikely to appear even for
generic values of the GUT  scale breaking, leave alone regions
where the coefficient function $F_I$ is itself large.

Even though they are ``captured'' at different orders in
perturbation theory the solutions obtained when $\hat  r_2$ is
assumed to be $\sim 1$ or $\sim\epsilon^2$ to begin with are very
close to the one displayed above which was obtained by assuming
that $\hat r_2\sim\epsilon$. In fact the final values of $\hat
r_2$ in these two cases are $.00185,.00225$. Some of the details
can be compared in Table I.

Focussing on the case I+2312(a) we find that using the value
$r_5'=-.333$  and the values of the ${\cal D}$  angles found for
the I+ solution :

\be \phi_{\cal D}=0.0525  \quad;~ \theta_{\cal D}= 0.43481 \quad;~
    \chi_{\cal D}= 0.02568\ee

 then gives, on using eqn.(\ref{DTF}) the following values for the eigenvalues of $\hat n$

\be \hat n_1 = 113.8 \quad;~ n_2 =20.045   \quad;~ n_3 =
1.97\times 10^{-5}  \ee

 Which clearly  shows the  required 100-300 fold enhancement relative
 to the BM fitting scenario in
  the overall scale of the Type I
 seesaw masses via an enhancement of  $\hat n$  that motivates our
 scenario.
  However the PMNS angles  are found to be :

  \be
\sin^2 \theta_{12}^{PMNS} =.073 \quad;~ \sin^2\theta_{23}^{PMNS} =
.77 \quad;~\theta_{13}=.176 \ee

Although one might hope to improve the value of the 23 and 13
angles towards more realistic values it is clear that  raising the
12 sector mixing from such a small value  will be difficult if the
fit is indeed stable as we propose it is. A final conclusion must
await a detailed survey.  However recall that even in the
$\mathbf{10-\oot}$ scenario  no Type I solution  was found
\cite{matsu0,babmacesanu} till CP violation was included in the
charged fermion sector. Results of the lepton sector parameters
for some for some of the other distinct cases are given for
comparison in Table 2.

 \begin{table}[t]
\begin{center}
\begin{tabular}{|c||c|c|c|c|c|c|}
\hline &I+&II+&III+&I-&II-&III-\\  \hline
N&15&15&9&18&20&11\\\hline $\delta$&1&0&2&1&0&$-1$\\
\hline
$ {\hat r}_2 $&.255&.225&.185& 1.21& 1.065&- 5.07\\
  $ r_6$&.01867&.01867&.018667&.01886&.01886&.018913\\
 $ r_7$&.0233&.0233&.0233&-.0234&-.0234&-.0233\\
$ \hat{h}_1$&.518 &.518&.518&.519&.519&.519\\
$\hat{h}_2$&.0277&.0277&.0277&.02689&.02687&.02651\\
$10^5 \hat{h}_3$&$1.92 $&$1.99 $&$2.05 $&$16 $&$16.55$&$24.5$\\
$ \hat{g} $&$\pm.122,0$&$\pm.122,0$&$\pm.122,0$&$\pm.120,0$&$\pm.120,0$&$\pm.120,0$\\
$10^4\hat{f}_1$&  1.33 &  1.17&
 .961 & 5.39 & 4.705 & 16.67  \\
$10^4 \hat{f}_2$&1.11&0.98&0.80&4.70&4.10&14.85\\
 $10^4 \hat{f}_3 $&0.171&0.150&0.12&0.76&0.66&1.60\\
 $\chi_{m^2}$&0.126&0.148&0.128&0.06&1.56&0.255\\
$\chi_{CKM}$&0.016&0.008&0.0132&1.641&0.075&0.051\\
  \hline
\end{tabular}
\end{center}
\caption{\label{tab1}\em Charged fermion fitting parameters and
Yukawa eigenvalues for 6 sample fits. N is the order of
perturbation theory at which the fit was ``captured'' while the
exponent $\delta$ specifies the initial assumption $\hat r_2\sim
\epsilon^\delta .$}  To the accuracy displayed the value of ${\hat
r}_1=15.27$ for every solution.
\end{table}

\subsection{(-) Type fits }

 For the branch defined by imposing
 $\chi_d^0=\chi_d^0=-\chi_l^0=-\chi_{\cal D}^0$ on  the  perturbative
 iteration,
  the fits we obtain are, at $\chi^2> 1.6 $,  somewhat poorer than
the ones found for the (+) case. Moreover even such fits first
occur at quite high order in the perturbation expansion and the
differences between the final $\hat r_2$ values are somewhat
greater. The acceptable cases with final values $\tilde r_2\sim
.01$ are shown in Table 1. as cases I-,II- along with another fit
III- where an initial value $\hat r_2\sim \epsilon^{-1}\sim $
leads to a distinct solution with $\tilde r_2\sim .05 $. This type
of solution has $\hat n$ eigenvalues $\{4.3,.75,.012\}$ and shows
clearly that $\hat f <.01 \epsilon^3$ is necessary to raise the
value of $\hat n$ enough to overcome the difficulty uncovered for
the MSGUT\cite{gmblm,blmdoom}. The rest of the procedure is
identical to the  $(+)$ case and the relevant results are
collected in Tables 1 and 2.

For the readers convenience we give the values of  the Yukawa
couplings, coefficients $r_i$ for one of the (-) cases. Namely
I-2312a.

 \bea {\hat r}_1 &=&   15.27   \quad ; \quad
          {\hat r}_2 = 1.21  \quad ; \quad r_6 =.01886\quad ; \quad
          r_7 =-.02340 \nnu
          \nnu
           {\hat{h}}  &=&
\left(\begin{array}{ccc}-0.000149706&-0.00054677&-0.014659\\-0.00054677&
    0.00927299&-0.134398\\-0.014659&-0.134398&0.48307
   \end{array} \right)\nnu
 \nnu{\hat{f}}  &=&
\left(\begin{array}{ccc}0.0000521467&0.000139&-0.00044\\0.000139&-0.000039&
    0.000182\\-0.00044&0.000182&0.000133
  \end{array} \right) \\
\nnu{\hat{g}}  &=& \left(\begin{array}{ccc}0 & 0.00116 & -0.01753 \\
 -0.00116 &0 &-0.1191  \\
 0.01753& 0.1191&0
  \end{array} \right)\nnu
\phi_{\cal D}&=& -0.04057 \quad ; \quad \theta_{\cal
D}=-0.134924\quad ; \quad \chi_{\cal D}=-0.57541\nonumber
  \eea

\bea M_U &=& \{95.148,0.2105,0.00077\}\nnu
 M_D &=&\{1.5835,0.0298017,0.0014962 \}\nnu
M_l&=& \{1.629,0.0753,0.000356\}\eea

  \begin{table}[t]
\begin{center}
\begin{tabular}{|c||c|c|c|c| }
\hline &I+2321a&I+2321b&I+2123a&I+2123c \\  \hline
 Hierarchy&$\tau>\mu>e$&$\tau>\mu>e$& $e>\mu>\tau$&
   $e>\mu>\tau$  \\
$r_5'$&-.333&.483&-.144&.302\\
  $ \hat{n}_1$& 113.83&104.26&54.38&51.95\\
$\hat{n}_2$&20.045&18.36&52.53&51.14\\
$\hat{n}_3$&$1.97\times 10^{-5}$&$1.89\times 10^{-5}$&$1.78\times
10^{-5}$&$1.774\times 10^{-5}$  \\
$\sin^2 2\theta_{23}$& .768 & .553 & .0726&.077\\
$\theta_{13}$&.176&.117&1.347&1.35\\
$\sin^2 2\theta_{12}$&.073&.130&.143&.753\\
  \hline
\hline
 &I-2312a&I-2312b&I-1223a &I-1223c \\
\hline
Hierarchy&$\tau>\mu>e$&$\tau>\mu>e$& $e>\mu>\tau$ & $e>\mu>\tau$     \\
$r_5'$&-.258&.698&-.073& .569\\
 $ \hat{n}_1$& 32.81&20.96&14.99& 11.12\\
$\hat{n}_2$&5.78&3.69&14.75& 10.95\\
$\hat{n}_3$&$4.65\times 10^{-4}$&$3.36\times 10^{-4}$&$4.53\times
10^{-4}$&$3.76\times 10^{-4}$   \\
$\sin^2 2\theta_{23}$& 0.886 & 0.447 &0.192&0.175\\
$\theta_{13}$&0.216&0.159&1.261&1.282\\
$\sin^2 2\theta_{12}$&0.209&0.206&0.494&0.772\\
  \hline\hline
\end{tabular}
\end{center}
\caption{\label{tab2}\em Values of neutrino mass and mixing
parameters determined assuming central value for the ratio of
solar to atmospheric mass splitting. }
\end{table}

\section{Discussion, Conclusions and Outlook}

In this paper we have reported progress in accomplishing our
program\cite{ag2,gmblm,blmdoom,nmsgutI}  for a completely
realistic fit of all known charged fermion and neutrino mass data
using the mass relations and RG evolution common to any SO(10)
Susy GUT with a $\mathbf{{10-120-\oot}}$  FM Higgs system fermion.
Specifically, we have shown that in the quasi realistic 3
generation but CP preserving(real) case we are able to obtain
accurate charged fermion fits, neutrino mass parameters and a PMNS
mixing pattern that can be large at least in the 23 sector and
small in the 13 sector. The remaining deficiencies, namely the
absence of simultaneous large mixing in the 12 sector and the fit
of the MSSM CKM CP phase in the first quadrant can presumably be
remedied in the complex 3 generation case, in close analogy with
the BM program where a succesful Type I fit could be
found\cite{matsu0,babmacesanu} \emph{only} when  CP violation was
introduced.  For the case where the couplings are all real and CP
violation arises from the phases present in the coefficients $r_i$
the CP violation can be traced to the GUT scale symmetry breaking
and fine tuning\cite{abmsv,ag2,bmsv}  that defines the light Higgs
pair of the MSSM. In this case the EW doublet($[1,2,\pm1]$) mass
matrix formulae for the NMSGUT that we have already
furnished\cite{blmdoom} will permit the evaluation of the
feasibility of the generic fit for realizability in the parameter
space of the NMSGUT. The report on the investigation of these
questions will be given in \cite{nmsgutIII}.

  We   carried  a somewhat
novel expansion in a single   parameter $\epsilon$, in terms of
which the very strong ($\sim 10^{5}$) hierarchy visible in the
MSSM fermion data can be compactly formulated,  to high orders of
the perturbation expansion . The explicit and systematic
incorporation of the hierarchical structure into our fitting
procedure makes it well suited to verification of the viability of
the hierarchical fitting scenario we
proposed\cite{blmdoom,nmsgutI}. On the other hand the convergence
in the parameter $\epsilon$ in the three generation case(in sharp
contrast to the 23 generation toy model\cite{nmsgutI}) is only
oscillatory\cite{nmsgutII}. The 2 generation Toy model is
necessarily CP conserving in the charged fermion sector whereas
the true solution  in the 3 generation case
 is not. Thus the failure to converge may reflect
an obstruction to fitting the data to the absolute minimum when
the channels to reach it on the complex plane have been blocked by
our reality assumption. Our method is complementary to alternative
numerical procedures such as the ``downhill simplex''
method\cite{grimuslavu,bertnu} which it can verify as well as
provide initial search points for. Inasmuch  as it is only the
actual hierarchical mass data that must be fitted the true
solution must necessarily be ``close'' to the 23 generation toy
model caricature. Thus the ``missed solutions anxiety'' associated
with these methods, may find some palliation when the fitting
problem is considered in the light of the insight provided by
these two complementary methods jointly. It is only after the
successful completion of this program that we may, finally, be in
possession of a well defined Susy GUT compatible with all low
energy data as well as information on the embedding of the MSSM in
the MSGUT : which emerges as the most valuable corollary product
of the fitting procedure\cite{gmblm}. It is only then that we will
be able to enter the third phase of the GUT program in which the
exotic process ($\Delta B \neq 0$ etc ) predictions will finally
begin to be formulated in a sufficiently tight manner as to make
the comparison with data a falsifiability test rather than a
hopeful check on a lottery bet.

\section{Acknowledgements}

I am grateful to Goran Senjanovic for unfailing enthusiasm and
hospitality and  the High Energy theory group of ICTP, Trieste for
hospitality. I also thank W. Grimus  for an invitation to  deliver
a seminar on this work at the Institut f$\ddot{u}$r Theoretische
Physik in Vienna in June 2006 and for discussion.

\section{Note Added}
On 18 July 2006, as this paper was being readied for submission to
the archive, we received \cite{grimusnu} in which a successful fit
to the complete fermion data including CP violation,for $\tan
\beta \sim 10$, with spontaneous CP violation,and with an
additional $Z_2$ symmetry imposed to further  reduce the number of
degrees of freedom,  is reported. The results are consonant  the
scenario proposed in \cite{blmdoom,nmsgutI} and the results of
this paper.

 \end{document}